\documentclass[journal]{IEEEtran}
\usepackage{amsmath,graphicx}
\usepackage{forloop,algpseudocode,algorithm,algorithmicx}
\usepackage{amsfonts}
\usepackage{multirow}
\ifCLASSINFOpdf
\else
   \usepackage[dvips]{graphicx}
\fi
\usepackage{url}

\usepackage{graphicx}
\usepackage[noadjust]{cite}

\begin{document}

\title{Multi-image Super-resolution via Quality Map Associated Attention Network}

\author{Minji Lee, \IEEEmembership{Student Member, IEEE}

\thanks{Minji Lee is with the School of Computing, KAIST, Daejeon, South Korea (e-mail: haewon\_lee@kaist.ac.kr).}}

\maketitle
\begin{abstract}
Multi-image super-resolution, which aims to fuse and restore a high-resolution image from multiple images at the same location, is crucial for utilizing satellite images. The satellite images are often occluded by atmospheric disturbances such as clouds, and the position of the disturbances varies by the images. Many radiometric and geometric approaches are proposed to detect atmospheric disturbances. Still, the utilization of detection results, \emph{i.e.}, quality maps in deep learning was limited to pre-processing or computation of loss. In this paper, we present a quality map-associated attention network (QA-Net), an architecture that fully incorporates QMs into a deep learning scheme for the first time. Our proposed attention modules process QMs alongside the low-resolution images and utilize the QM features to distinguish the disturbances and attend to image features. As a result, QA-Net has achieved state-of-the-art results in the PROBA-V dataset.
\end{abstract}


\IEEEpeerreviewmaketitle

\section{Introduction}
\IEEEPARstart{S}{uper-resolution} (SR) aims to reconstruct a high-resolution (HR) image from either single or multiple low-resolution (LR) images. Due to the expensive cost of high-precision sensors, super-resolution is an extensively studied subject in remote sensing. Recent single-image super-resolution methods \cite{1,6,7,8,9} have shown outstanding performances, utilizing deep neural networks to represent the mapping function from the LR image to the target HR image. However, the high-frequency details in a single image are limited, and the lost details are hard to be recovered.

Multi-image super-resolution (MISR) exploits multiple LR images to generate an HR image. MISR has gained massive interest due to its use in Earth observation satellite images, as the satellite can capture images of the same location over time. Note that MISR differs from video super-resolution (VSR), although they both employ multiple LR frames. VSR aims to reconstruct a high-resolution video of a changing scene, and the frame sequence has a very short time interval. The input images in MISR do not have an explicit temporal relation and capture the scenery of the same location.

\begin{figure}[t]
\label{fig:1}
\centerline{\includegraphics[width=\columnwidth]{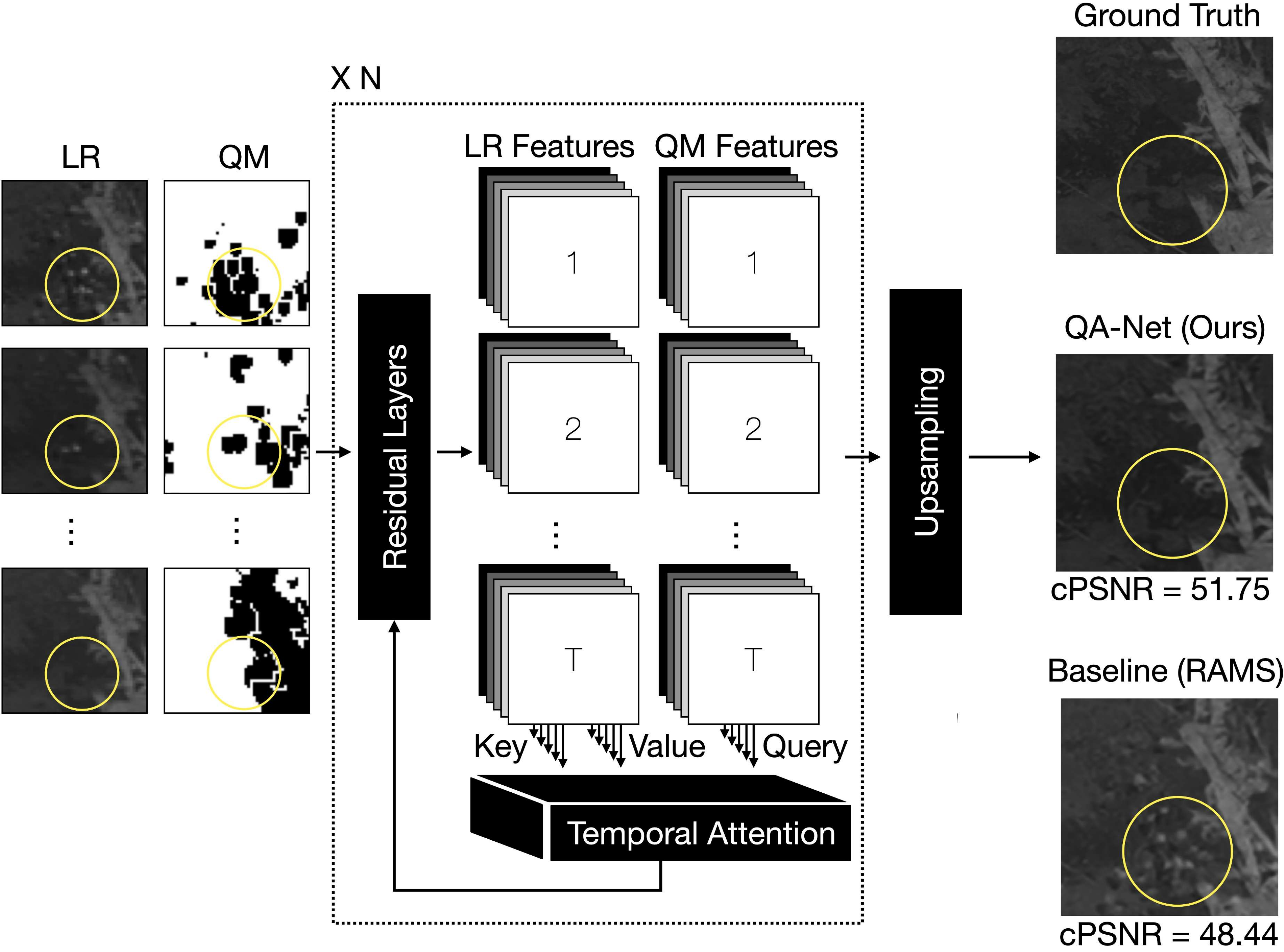}}
\caption{Illustration of the proposed method. The quality map features are repetitively exploited to attend the low-resolution image features during feature extraction and fusion.}
\end{figure}

There are two main concerns about the MISR of satellite imagery: (i) misalignments such as rotation and shift between the LR images, and (ii) atmospheric disturbances such as clouds and shadows degrading the quality of the images. Previous works \cite{2, 3, 5, 11, 13} have mainly concentrated on the alignment and the fusion of the LR images, and the local deviation of image quality due to atmospheric disturbances has been overlooked. Even though comparing images is crucial to distinguish atmospheric disturbances, many MISR methods compute each image separately.

We argue that it is necessary to discriminate the region hindered by the disturbances and concentrate on the reliable features to generate a better super-resolved image. It can be accomplished by encouraging communication among the images during feature extraction and effectively using the \emph{quality maps}. A quality map indicates the area of the remotely sensed image corrupted by the atmospheric disturbances and can be generated using algorithms \cite{18} comparing the retrieved reflectances of different bands. QMs provide helpful information on the reliability of the features, but previous deep learning-based MISR methods only used QMs for image selection, registration, and as the mask to compute loss.

\begin{figure*}[t]
\centerline{\includegraphics[width=\linewidth]{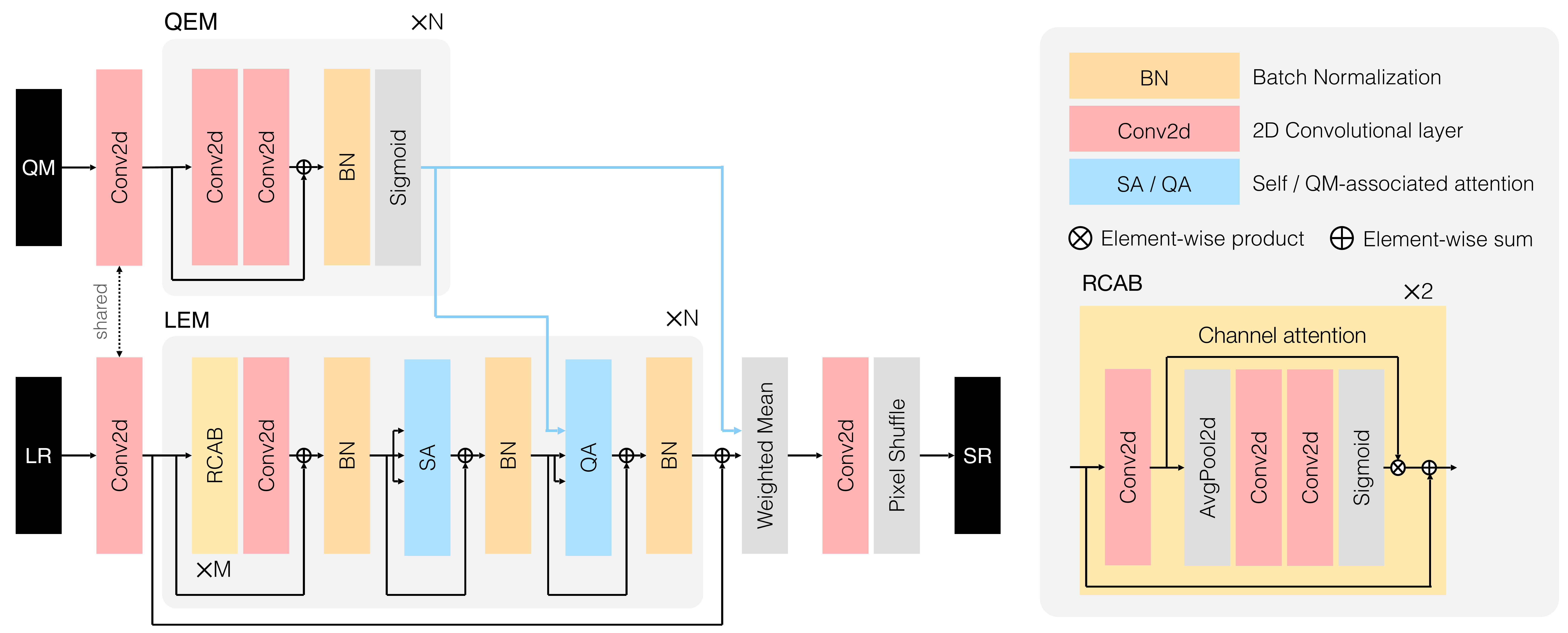}}
\centering
\caption{Model architecture of QA-Net. We omitted ReLU activation functions for simplicity. LR image encoding modules (LEM) and QM encoding modules (QEM) are repeated and for each iteration the output of QEM are used in the quality map-associated module of LEM.} 
\end{figure*}

In this paper, we explore the usage of QMs further and propose the quality-map associated attention network (QA-Net). Our model introduces QMs in both the feature extraction of the LR images and their fusion, as depicted in Fig. 1. QA-Net simultaneously processes the LR image and QM pairs in separate repeated modules, and the processed QM features are utilized to attend the features from the LR images for every iteration. With the guidance of QMs, QA-Net can locally concentrate on the features of the less-disturbed images and fully exploit the advantages of utilizing multiple LR images. As a result, our QA-Net achieves state-of-the-art performance on the PROBA-V dataset.

\section{Related Works}
MISR aims to effectively align and fuse multiple images captured in the same location to reconstruct a higher resolution image. In MISR, LR images are first registered, \emph{i.e.,} aligned against sub-pixel shifts and rotations. Some methods \cite{2, 12} select or generate reference images and register LR images to the reference image using convolutional layers. Others \cite{3, 4, 13} register images in the Fourier domain using masked normalized cross-correlation (NCC), \emph{i.e.} masked FFT NCC \cite{14}, and optical flow estimation. 

As well to the aforementioned registration methods, many deep learning-based fusion methods are proposed. HighRes-net \cite{2}  fuses images using recursive convolutional layers, and MISR-GRU \cite{11} fuses images using a recurrent neural network. However, such methods that process images in the separate convolutional or recurrent module are incapable of learning relations between images. Recently, TR-MISR \cite{13} used a pixel-wise fusion method to learn the relation between features of images.

\section{Methodology}
This section presents QA-Net, a novel end-to-end neural network for multi-image super-resolution. QA-Net consists of five parts: shallow feature extraction, LR image encoding modules (LEMs), QM encoding modules (QEMs), weighted mean, and upsampling. Let the low-resolution images of a scene and their quality maps, $\text{LR}_{i}$, $\text{QM}_i$, respectively, where $i = 1, 2, \cdots, T$ and $T$ is the number of the images. $\text{QM}_i$ is a binary map where the value of 0 indicates the disturbances such as clouds, shadows, ice, or water, and the value of 1, \emph{i.e.}, clear pixel, indicates the area of interest. QA-Net uses a shared convolutional layer $\text{SF}_\theta$ to extract a shallow feature from each LR image and QM:
\begin{eqnarray}
L^0_i &= \text{SF}_\theta (\text{LR}_i), \quad Q^0_i =  \text{SF}_\theta (\text{QM}_i).
\end{eqnarray}
$\{L^0_i\}^{T}_{i=1}, \{Q^0_i\}^{T}_{i=1}$ pass through $\text{LEM}_{j}$ and $\text{QEM}_{j}$ respectively, where $j = 1, 2,\cdots, N$ and $N$ denotes the number of encoding modules:
\begin{align}
\{L^j_i\}^{T}_{i=1} & = \text{LEM}_j(\{L^{j-1}_i\}^{T}_{i=1}, \{Q^{j-1}_i\}^{T}_{i=1}),\\
\{Q^j_i\}^{T}_{i=1} & = \text{QEM}_j(\{Q^j_i\}^{T}_{i=1}).
\end{align}
After the encoding, the shallow feature $L_i^0$ are added to the last LR features $L_i^N$ and are accumulated with a weighted mean using the QM features as the weight, resulting in a fused feature $f$ for every pixel $(x, y)$:
\begin{eqnarray}
f(x, y) = \dfrac{ \sum_{i=1}^T (L_i^0 (x,y) + L_i^N(x,y))\cdot Q_i^N (x,y)}{\sum_{i=1}^T Q_i^N (x, y)}
\end{eqnarray}
Finally, the super-resolved image SR is achieved via upsampling $f$ with a 2D convolutional layer and pixel shuffle.

\subsection{LR image encoding modules (LEMs)}
LEM mainly consists of self-attention \cite{17} module (SA), QM-associated attention module (QA), and multiple residual channel attention blocks (RCABs) \cite{9}. SA learns the relation between LR features, while the QA computes the attention map based on features of QM and LR to let the network focus more on LR features with better quality. 

LR features $\{L^j_i\}_{i=1}^T$ and QM features $\{Q^j_i\}_{i=1}^T$ are divided to $H\times W$ pixels $l_{(x,y)}=\{l_{i, (x,y)}\}_{i=1}^T, q_{(x,y)}=\{q_{i,(x,y)}\}_{i=1}^T \in \mathbb{R}^{T\times C}$ where $C$ denotes the number of channels. Here, we omit $j$ since attention modules are repeated for every LEM. Self-attention compares $l_{(x,y)}$ along the image dimension $T$ to attend clear feature using learnable parameter matrices $W_Q, W_K\in \mathbb{R}^{C\times T}$, resulting a self-attention map $\text{SAM} \in \mathbb{R}^{T \times T}$:
\begin{eqnarray}
\text{SAM}(l_{(x,y)}) = \text{Softmax} \bigg(\dfrac{l_{(x,y)}\; W_Q \; W_K^{T}\; l_{(x,y)}^T} {\sqrt{T}}\bigg).
\end{eqnarray}
SAM is multiplied by $l_{(x,y)}$ to weigh the clear features. Attended features are further projected using learnable parameter matrices $W_V \in \mathbb{R}^{C\times T}, W_0 \in \mathbb{R}^{T\times C}$:
\begin{eqnarray}
\text{SA}(l_{(x,y)})=\text{SAM}(l_{(x,y)})\; l_{(x,y)} W_V W_0.
\end{eqnarray}
Similarly, QA compares $l_{(x,y)}, q_{(x,y)}$ along the image dimension $T$ using learnable parameter matrices $V_Q, V_K\in \mathbb{R}^{C\times T}$, resulting a QM-associated attention map $\text{QAM} \in \mathbb{R}^{T \times T}$:
\begin{eqnarray}
\text{QAM}(l_{(x,y)},q_{(x,y)}) = \text{Softmax} \bigg(\dfrac{q_{(x,y)}\; V_Q \; V_K^{T}\; l_{(x,y)}^T} {\sqrt{T}}\bigg).
\end{eqnarray}
By multiplying QAM by $l_{(x,y)}$, the LR feature with a better corresponding QM feature is weighted. Then QM-associated attention is completed by projecting learnable parameter matrices $V_V \in \mathbb{R}^{C\times T}, V_0 \in \mathbb{R}^{T\times C}$:
\begin{eqnarray}
\text{QA}(l_{(x,y)},q_{(x,y)})=\text{QAM}(l_{(x,y)},q_{(x,y)})\; l_{(x,y)} V_V V_0.
\end{eqnarray}
Altogether, repeated LEMs allow the network to compare features of multiple images during the feature extraction process. 

\begin{algorithm}[t!]
\caption{Quality-map associated image selection}
\label{alg:sel}
\begin{algorithmic}
\Require $\text{Corresponding QMs } \text{QM}=\{\text{QM}_1, \cdots, \text{QM}_N \}$
\Require $\text{Number of input images }T$
\Require $\text{Number of pixels to compare }p$
\State $\mathcal{D} \gets \{ \}, \mathcal{Q} \gets \{\}$
\For{$t \in [0, \text{Number of iterations}]$} 
    \State {$d \gets$ Set of randomly sampled $T$ indices}
    \State {$A \gets \displaystyle \sum_{i \in d} \text{QM}_i$}
    \State {$\mathcal{D} \gets \mathcal{D} \cup d$}
    \State {$\mathcal{Q} \gets \mathcal{Q} \cup \displaystyle \sum (p$ minimum values of $A$)}
\EndFor
\State {$m \gets \text{argmin} \mathcal{Q} $}
\\
\Return {$\mathcal{D}_m$}
\end{algorithmic}
\end{algorithm}

\subsection{QM encoding modules (QEMs)}
Compared to LRs, QMs are processed using a relatively shallow module composed of two convolutional layers with a skip connection. Since the features are used as the query of the attention module, we used the sigmoid activation function.

\subsection{Image selection}
Since the number of LR images can vary by location, $T$ LR images are selected according to the Algorithm \ref{alg:sel} in test time. For training, $T$ images are randomly selected according to the probability proportional to the number of clear pixels in the corresponding QM. Quality-map-associated image selection algorithm leads to uniform distribution of clear pixels throughout the images, thus leveraging the attention process. 

\begin{table}[t!]
\small
\setlength{\tabcolsep}{3pt}
\centering
\caption{Quantitative comparison between different models on the PROBA-V validation set.} 
\label{tab:2}
\bgroup
\def\arraystretch{1.2}
\begin{tabular}{|c|c|c|c|c|}
  \hline
  \multirow{2}{*}{Model} & \multicolumn{2}{c|}{RED} & \multicolumn{2}{c|}{NIR} \\
  \cline{2-5}
   & cPSNR (dB) & cSSIM & cPSNR (dB) & cSSIM \\ 
  \hline
  Bicubic & 47.34 & 0.9846 & 45.44 & 0.9767 \\
  IBP \cite{10} & 48.21 & 0.9865 & 45.96 & 0.9796 \\
  RCAN \cite{9} & 48.22 & 0.9870 & 45.66 & 0.9798 \\
  HighRes-net \cite{2} & 49.75 & 0.9904 & 47.55 & 0.9855 \\
  DeepSUM \cite{12} & 50.00 & 0.9908 & 47.84 & 0.9858 \\
  MISR-GRU \cite{11} & 50.11 & 0.9910 & 47.88 & 0.9861 \\
  RAMS \cite{3} & 50.17 & 0.9913 & 48.23 & 0.9875 \\
  TR-MISR \cite{13} & 50.67 & 0.9921 & 48.54 & 0.9882 \\
  PIUnet \cite{4} & 50.62 & 0.9921 & 48.72 & 0.9883\\ 
  \textbf{QA-Net (ours)} & \textbf{50.82} & \textbf{0.9976} & \textbf{48.81} & \textbf{0.9964}  \\
  \hline
\end{tabular}
\egroup
\end{table}

\section{Experimental Settings}
QA-Net is trained and tested on the PROBA-V dataset \cite{19}, which is composed of Top-Of-Atmosphere reflectances for red visible (RED) and Near IR (NIR) spectral bands at 300-m (128$\times$128 grayscale image) and 100-m (384$\times$384 grayscale image) resolution. We aim to reconstruct 100-m resolution (HR) images from multiple 300-m resolution (LR) images.

\subsection{Implementation details}
LR images are registered using masked FFT NCC \cite{14} and cropped to 32$\times$32 patches before the training process. We set the number of RCABs in LEM $M=8$, LEMs and QEMs $N=12$, input LR and QM pairs $T=9$, and the pixels to compare $p=50$. The kernel size of the 2D convolution is set to 3$\times$3. We used the Adam optimizer with the initial learning rate set to 1e-4 and the weight decay set to 1e-5. The network is trained 520 epochs and the learning rate is multiplied by 0.8 at 120, 220, 300, 360, 400, 440, and 480 epochs.

\subsection{Evaluation metric}
Assuming that pixel intensities are real numbers ranging from 0 to 1, the quality of a super-resolved image is evaluated with the clear peak signal-to-noise ratio (cPSNR) via the following calculations. First, we calculate the bias in brightness:
\begin{eqnarray}
b = \frac{1}{|P|} \sum_{(x,y) \in P}(\text{HR}(x,y)-\text{SR}(x,y)),
\end{eqnarray}
where $P$ represents a set of clear pixels in a high-resolution image. Then we calculate the clear mean squared error (cMSE) and cPSNR as follows.
\begin{align}
\text{cMSE} &=  \frac{1}{|P|}\sum_{(x,y) \in P} (\text{HR}(x,y)-\text{SR}(x,y)+b)^2,\\
\text{cPSNR} &= -10 \cdot \log_{10} (\text{cMSE}).
\end{align}
We used cPSNR between the ground truth and the output as a loss function. Similarly, we define clear structural similarity index measure (cSSIM) as SSIM calculated on clear pixels. In the test phase, the super-resolved images are cropped by a 3-pixel border, and the highest cPSNR among corresponding patches around the center of the ground-truth images is used as the score to compensate for pixel shifts.

\begin{figure*}[t]
    \centerline{\includegraphics[width=0.84\linewidth]{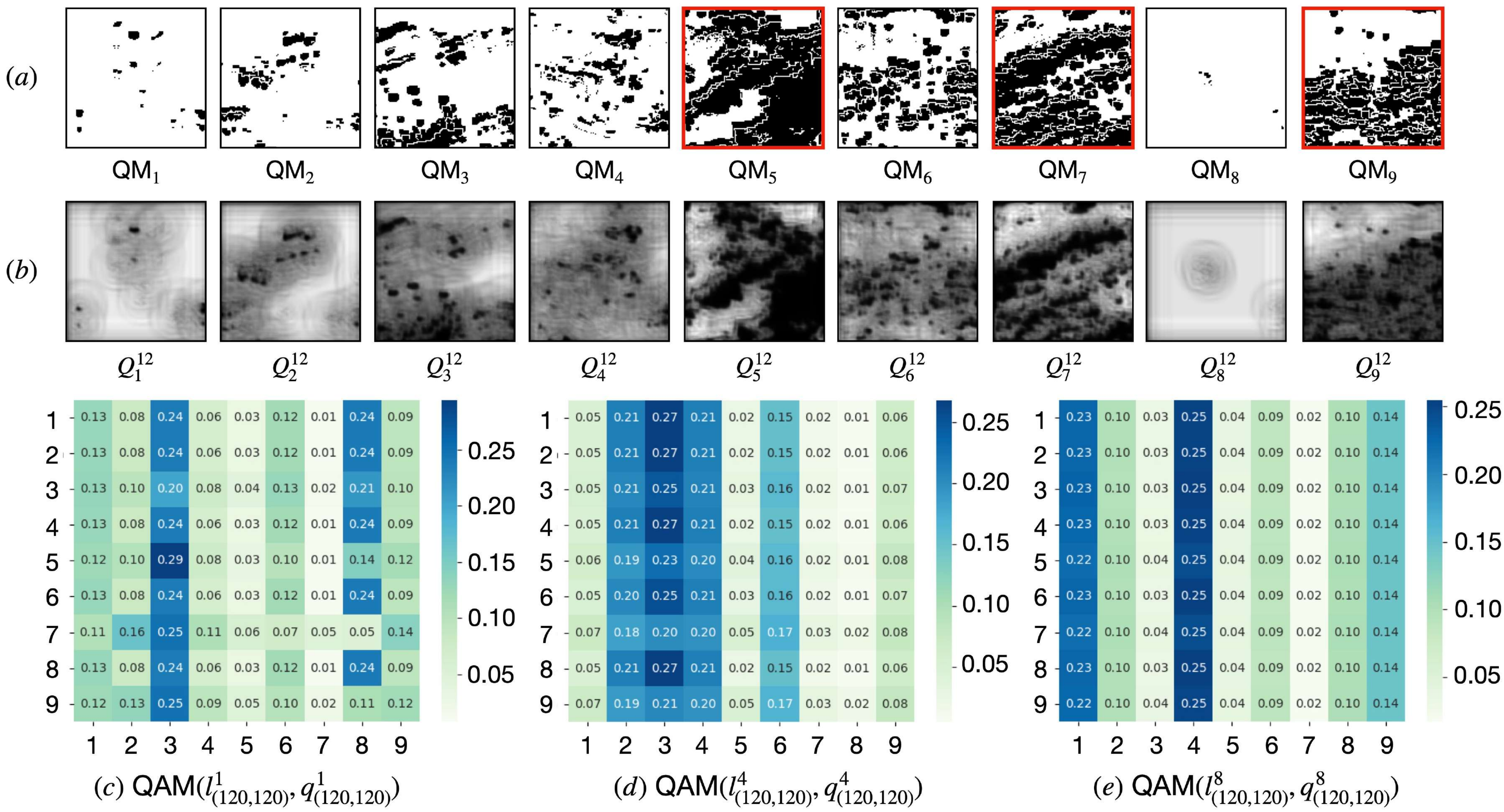}}
    \caption{QEM output and attention map visualization on NIR imgset1121. (a) Input QMs. Red boundaries denote that the pixel (120, 120) is disturbed. (b) Normalized outputs of the last QEM. (c)-(e) QM-associated attention map (QAM) of the pixel (120, 120) of first, fourth, and eighth LEMs, respectively.}
    \label{fig:qem}
\end{figure*}

\section{Results}
\subsection{Quantitative comparisons}
Table \ref{tab:2} shows the comparison of cPSNR and cSSIM with several methods on the PROBA-V validation set. Since the ground truth of the test set is not publicly available, we compared models on the validation set. Our proposed method outperforms the previous methods by 0.20 dB in the RED spectral band and 0.09dB in the NIR spectral band in PSNR. Also, our method achieves the best cSSIM results in both spectral bands with a significant performance gap.

\subsection{Qualitative comparisons}
As shown in the example of Fig. 4, the region in the yellow box is captured clearly in only a few LR images, \emph{e.g.}, LR000. Previous methods cannot recover the region accurately since the images are hindered by atmospheric disturbances in other LR images, \emph{e.g.}, LR003. On the contrary, our QA-Net recovered the region properly with the aid of QMs and communication among LR images. Our quality-map-associated attention allows the network to unveil the disturbed feature of an LR image using the clear feature from another LR image.

\section{Discussion}
Fig. 3 (a)-(b) shows the QMs and normalized feature of their last QEM output, respectively. As the convolution operations in QEMs increase the receptive field and the number of channels for each pixel, QEM outputs have richer information about the atmospheric disturbances than the binary values of QMs.
Fig \ref{fig:qem} (c)-(e) shows the QM-associated attention map (QAM) of the pixel (120, 120). Attention is computed through: 
\begin{eqnarray}
    \{\text{QAM}_{k,1} \; l_{1, (120, 120)} + \cdots + \text{QAM}_{k,9} \;l_{9, (120, 120)}\}^9_{k=1}.
\end{eqnarray}
The most weighted vectors are $l_{1, (120, 120)}, l_{8, (120, 120)}$ in the first QAM, $l_{2, (120, 120)}, l_{3, (120, 120)}, l_{4, (120, 120)}$ in the fourth QAM, and $l_{1, (120, 120)}, l_{4, (120, 120)}$ in the eighth QAM. $l_{5, (120, 120)}, l_{7, (120, 120)}, l_{9, (120, 120)}$, is disturbed heavily (marked with red box in Fig. \ref{fig:qem} (a)) so are never highly weighted. However, the highly weighted vectors differ by every LEM, which implies that each LEM focuses on different features, similar to the convolutional filters in CNN.

\section{Conclusion}
In this paper, we introduced a novel architecture that can distinguish atmospheric disturbances in satellite imagery named QA-Net. QA-Net is the first multi-image super-resolution method to exploit quality maps to attend low-resolution images in feature extraction and fusion processes. Quantitative and qualitative evaluations on the PROBA-V dataset show the outstanding performance of our method, thus highlighting the importance of utilizing quality maps in deep learning-based MISR, and the comparison and attention across images.

\begin{figure}[t]
    \centerline{\includegraphics[width=0.8\columnwidth]{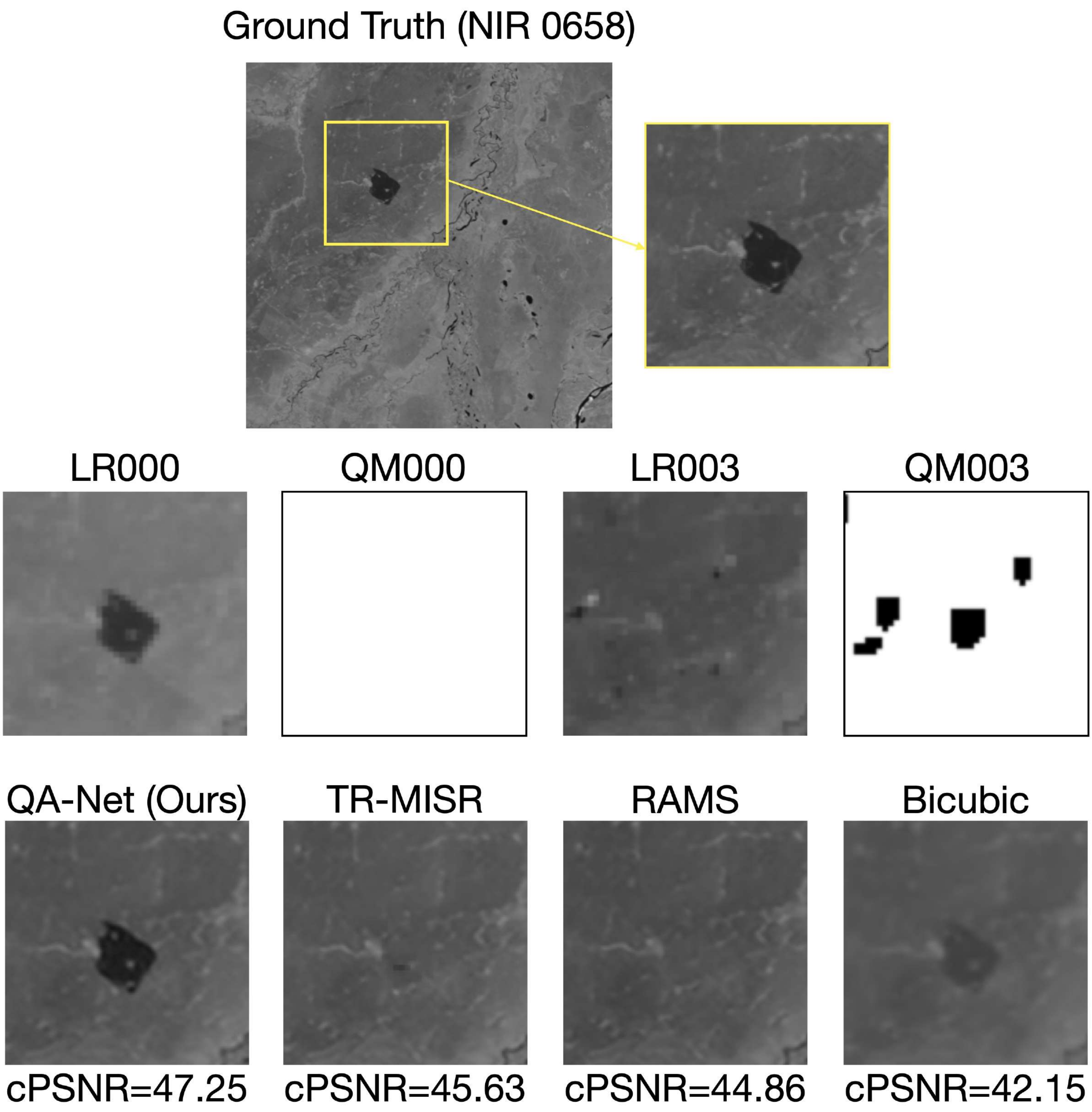}}
    \caption{Visual comparison between different models on NIR imgset0658.}
    \label{fig:3}
\end{figure}

\bibliography{main}
\bibliographystyle{IEEEtran}

\end{document}